# Free-breathing Pulmonary Relaxometry at 0.55T

**Authors:** Pavlos Panos[1,2], Oliver Bieri[1,2], Maurice Pradella[3], Katrin E. Hostettler[4], Grzegorz Bauman[1,2]

**Author Affiliations:**

*[1]Department of Biomedical Engineering, University of Basel, Allschwil, Switzerland*

*[2]Department of Radiology, Division of Radiological Physics, University Hospital Basel, Basel, Switzerland*

*[3]Department of Radiology, Division of Cardiothoracic Imaging, University Hospital Basel, University of Basel, Basel, Switzerland*

*[4]Clinics of Respiratory Medicine, University Hospital Basel, Basel, Switzerland*

**Corresponding author:**

Pavlos Panos

Division of Radiological Physics

Department of Radiology

University Hospital Basel

Petersgraben 4

CH-4031 Basel

Switzerland

E-mail: pavlos.panos@unibas.ch

## ABSTRACT

**Purpose:**

To evaluate the feasibility of an integrated, free-breathing workflow for automated 2D pulmonary relaxometry (T1, T2) at 0.55T.

**Methods:**

A 2D inversion recovery ultra-fast balanced steady-state free precession (IR-uf-bSSFP) sequence was adapted to achieve high-temporal sampling of the transient phase at 0.55T. The technique was validated in a phantom and tested in eight healthy volunteers as well as one patient. A fully automated pipeline was developed, featuring multi-contrast registration for motion correction and deep learning-based lung segmentation to enable voxel-wise nonlinear fitting for T1 and T2 map generation.

**Results:**

Phantom results were in close agreement with reference scans. In-vivo, the proposed free-breathing framework effectively mitigated respiratory motion, yielding quantitative maps in close agreement with breath-hold references. Healthy lung parenchyma relaxation times were T1 = (930$\pm$40)ms and T2 = (90$\pm$8)ms. In a patient case, the method successfully distinguished a solid lung mass from healthy parenchyma, with the lesion showing elevated T1 (960ms vs 810ms in the surrounding parenchyma).

**Conclusions:**

Simultaneous free-breathing T1 and T2 mapping of the lung is feasible at 0.55T using a fully automated pipeline. By eliminating breath-holds and external gating, this approach improves patient compliance and potentially facilitates the use of quantitative lung MRI in routine clinical practice.

# 1. INTRODUCTION

Standard clinical magnetic resonance imaging (MRI) traditionally relies on qualitative morphological assessment; however, it also offers the potential for quantitative tissue characterization [1–5]. This can be achieved by generating spatially resolved maps of physical parameters, such as the longitudinal T1 and transverse T2 relaxation times. While these metrics could serve as non-invasive biomarkers for monitoring therapeutic responses or assessing disease severity, they remain underexplored, partly due to current technical limitations [3,6].

MRI has gained significant traction in thoracic imaging due to its excellent soft-tissue contrast and ionizing radiation-free nature [7–9], yet translating advanced quantitative methods to the lung remains challenging [10,11]. The primary difficulties are rooted in the lung's intrinsic structure: the air-filled, sponge-like parenchyma has an inherently low proton density, and extensive air-tissue interfaces generate severe magnetic susceptibility artifacts [12]. This combination causes the MR signal to decay extremely rapidly, resulting in an effective transverse relaxation time (T2*) as short as 8−10 ms at 0.55T [13]. These challenges are further amplified by continuous cardiorespiratory motion, which degrades image quality and compromises quantitative precision.

To overcome these limitations, various imaging protocols have been adapted for pulmonary relaxometry. Initial success in T1 quantification was achieved using established sequences such as snapshot fast low-angle shot (FLASH) [14] or inversion recovery half-Fourier single-shot turbo spin-echo (IR-HASTE) [15]. More recent methods, such as radial k-space sampling with ultra-short echo time (UTE) [16] and magnetic resonance fingerprinting at 0.55T [17], have enabled the simultaneous multi-parametric quantification. While these approaches have proven valuable for characterizing diseases like emphysema and fibrosis [18,19], the role of T2 as an independent biomarker is less understood [20]. Furthermore, persistent challenges including breath-holding requirements, long acquisition times, and complex workflows continue to hinder routine clinical adoption.

Simultaneous T1 and T2 quantification can be achieved by analyzing the transient signal evolution of balanced steady-state free precession (bSSFP) following a magnetization preparation, such as inversion or saturation recovery pulse [21,22]. However, the pronounced sensitivity of bSSFP to off-resonances can critically affect relaxometry at high fields or in organs with large magnetic susceptibility gradients, such as the lung, particularly if not explicitly addressed.

The concept of simultaneous T1 and T2 quantification based on bSSFP signal evolution was first introduced by Schmitt et al. [22]. This approach was subsequently adapted for pulmonary imaging by Bauman et al. [23], who employed an ultra-fast bSSFP (uf-bSSFP) sequence variant to address the challenging structure of the lung tissue at 1.5T. By minimizing the repetition time (TR), the uf-bSSFP sequence broadens the spectral passband. This modification serves two critical functions: it mitigates signal loss from macroscopic field inhomogeneities, thereby reducing banding artifacts, and it more effectively refocuses local static intravoxel dephasing, which increases the characteristically low signal from the lung parenchyma.

While this technique enabled fast, simultaneous T1 and T2 mapping, its clinical utility was constrained by a reliance on breath-holding to mitigate respiratory motion and on ECG-triggering. These dependencies not only increase workflow complexity and scan time but also render the examination challenging or infeasible for many patients, including pediatric and elderly subjects, as well as those with severe respiratory conditions.

The goal of this work is to address these practical challenges by evaluating the feasibility of an integrated, free-breathing relaxometry workflow at 0.55T. Our approach eliminates the constraints of breath-holding and gating equipment by implementing a free-breathing acquisition strategy. To achieve this, we adapted the established IR-uf-bSSFP method [23,24] for use at a lower field strength of 0.55T, accounting for limited gradient performance, and developed an automated post-processing pipeline consisting of non-rigid motion correction and deep learning-based segmentation. This study describes the initial evaluation and feasibility of this integrated, free-breathing framework for pulmonary relaxometry.

## 2. METHODS

### 2.1 MR Pulse Sequence and Imaging Parameters

All experiments were performed on a whole-body 0.55T MR-scanner (MAGNETOM Free.Max, Siemens Healthineers, Forchheim, Germany), equipped with low-performance gradients (26 mT/m amplitude, 45 T/m/s slew rate). The integrated body coil was used for RF transmission and a combination of a 6-channel chest array coil and a 6-channel spine coil was used for signal reception. The study was approved by the local Institutional Ethics Committee, and written informed consent was obtained from eight healthy volunteers (mean age: 29 years; 4 female, 4 male) and from one patient (62 years old male with confirmed lung cancer and mediastinal lymph node metastasis).

We adapted a custom 2D pulse sequence based on an IR-uf-bSSFP design [23]. Following a non-selective adiabatic hyperbolic-secant inversion pulse, a continuous series of snapshot bSSFP images was acquired during the bSSFP's transient response toward steady-state. Key imaging parameters were as follows: TE=1.36 ms, TR=3.0 ms, nominal flip angle (α)=45°, FOV=475×475 $mm^2$, a 128×128 acquisition matrix (interpolated to 256×256 in post-processing), in-plane resolution of 3.7×3.7 $mm^2$ (interpolated to 1.9×1.9 $mm^2$), slice thickness of 14.0 mm, and a parallel imaging (GRAPPA) factor of 2. The single-image acquisition time (TA) was 192 ms. To ensure denser sampling of the signal recovery curve and mitigate the prolonged TR caused by gradient limitations at 0.55T, we employed a strategy using two separate inversion acquisition blocks. Imaging was repeated with an initial delay of half of the TA (96 ms), effectively doubling the sampling rate as compared to one single inversion acquisition block. A total of 24 inversion time points were acquired in two blocks of 12 (**Figure 1**).

The entire acquisition was performed during free-breathing, without the use of cardiac or respiratory triggering or gating. In contrast, respiratory motion was managed retrospectively using a dedicated multi-stage image registration. Furthermore, to address the inherent signal to noise ratio (SNR) challenges at low field strength and average out potential perfusion-related signal modulations, four measurements were acquired for each time point and subsequently averaged. Data acquisition was performed sequentially, completing all measurements for a given slice before moving to the next. To ensure sufficient longitudinal magnetization recovery, a delay of 500 ms was applied between consecutive inversion blocks. This protocol resulted in an acquisition duration of approximately 18 seconds per slice, yielding a total scan time of 3:04 minutes for 10 slices.

To evaluate the baseline thoracic field environment and verify the assumption of negligible off-resonance effects on the transient bSSFP signal evolution at 0.55T, B0 field maps were acquired in a subset of healthy volunteers (N = 3). Field mapping was performed according to a previously validated approach [25] to map the baseline field distribution within the lung parenchyma.

**2.2 Phantom Validation Acquisition**

To evaluate the quantitative accuracy of the adapted IR-uf-bSSFP sequence, a phantom was scanned at 0.55T. Reference T1 and T2 maps were generated using spin-echo sequences. The baseline T1 reference map was computed by fitting a series of inversion-recovery turbo spin-echo (IR-TSE) images acquired with a repetition time (TR) of 5100 ms, an echo time (TE) of 15 ms, an echo train length (ETL) of 17, a 96 × 96 acquisition matrix, and 13 distinct inversion times (TIs = 50, 100, 200, 350, 500, 700, 900, 1200, 1600, 2000, 2500, 3200, and 4000 ms). The baseline T2 reference map was generated using a turbo spin-echo (TSE) protocol utilizing a TR of 5000 ms, an ETL of 17, a 128 × 128 acquisition matrix, and a full train of 15 sequentially sampled echo times (TEs = 15, 29, 44, 59, 73, 88, 103, 117, 132, 147, 161, 176, 191, 205, and 220 ms).

**2.3 Automated Image Processing: Motion Correction**

To address the dual challenges of respiratory motion and the large contrast variations inherent to the IR-uf-bSSFP time series, we developed a fully automated, two-step image registration approach (**Figure 1**). This framework fundamentally relies on multiple acquisitions to provide a consistent mid-breathing reference target. By acquiring four repetitions per image contrast, the unguided respiratory cycle is sufficiently sampled to guarantee that at least one frame coincides the desired mid-respiratory phase.

Aligning each frame to this stable reference, rather than performing pairwise registration between frames, improves structural stability during the motion correction, despite the highly disparate contrasts.

The initial step corrects for respiratory motion among images acquired at the same inversion time point. Within each time-point-specific set, we identified a reference frame corresponding to the phase closest to the mid-respiratory cycle. This was achieved by extracting a respiratory signal from diaphragmatic motion, which was then smoothed and normalized to identify the mid-cycle phase. To ensure robust alignment, an edge-preserving filter was applied to each image to reduce noise while preserving anatomical detail [26]. Subsequently, a diffeomorphic 'demons' algorithm [27] performed non-rigid registration to warp each frame to the selected mid-respiratory reference.

Following the initial motion correction, the images within each time-point were averaged to produce a single mid-phase image with an enhanced SNR. The second step was to co-register this resulting series of averaged images. However, this task is much more challenging due to the large contrast differences between different inversion time points. For this purpose, we employed a symmetric diffeomorphic registration algorithm [28]. After normalizing image intensities to a consistent range, the algorithm aligned all images by deforming them into a shared intermediate space rather than a single reference image. This symmetric approach minimizes bias arising from the selection of the fixed target image, a common issue in traditional registration methods [29]. A cross-correlation metric was chosen to guide the registration due to its tolerance for non-linear intensity changes [30]. The entire process was accelerated by optimizing the algorithm's iterative pyramid scheme.

### 2.4 Automated Image Processing: Lung Segmentation

Following image co-registration, quantitative parameter fitting was spatially restricted to the lung parenchyma to reduce computational overhead. Because the blood pool within large pulmonary vessels exhibits fundamentally distinct relaxation properties than the thin, air-filled alveolar structures of the parenchyma, including major vascular elements would severely confound the averaged parenchymal relaxation metrics and mask subtle tissue-specific changes. To prevent this signal contamination, a two-stage pipeline combining two deep learning models was implemented. First, a whole-lung mask was generated using an established deep learning model adapted from Pusterla et al. [31]. This mask was then refined using a secondary, deep learning-based vessel segmentation step to identify and exclude the pulmonary vasculature, following the approach described by Panos et al. [32]. This targeted exclusion isolated the clean lung parenchymal tissue for subsequent voxel-wise T1 and T2 estimation.

### 2.5 Quantitative Mapping and Slice Profile Correction

Quantitative maps were generated by applying a voxel-wise non-linear fit to the signal time course, S(t), for all voxels within the lung mask. The signal recovery was modeled using a three-parameter equation describing the transient bSSFP signal after an inversion pulse [22]:

$$S(t) = S_{stst}\left(1 - INV \cdot e^{-\frac{t}{T1^*}}\right) \quad (1)$$

where $S_{stst}$ is the steady-state signal, INV is the inversion factor, and $T1^*$ is the apparent relaxation time. From these fitted parameters, the quantitative maps of longitudinal relaxation time (T1), and transverse relaxation time (T2) were calculated using the following relationships [22]:

$$T1 = T1^* \cdot \cos\left(\frac{\alpha}{2}\right) \cdot (INV - 1) \quad (2)$$

$$T2 = T1^* \cdot sin^2\left(\frac{\alpha}{2}\right) \cdot \left(1 - \frac{\cos\left(\frac{\alpha}{2}\right)}{INV - 1}\right)^{-1} \quad (3)$$

Note that Equations (1)-(3) refer to a single flip angle. To account for the flip angle dependency of these model equations, a correction for the non-ideal 2D slice profile was performed. While B1 field inhomogeneities can confound measurements by scaling the nominal flip angle, the B1 field at 0.55T is sufficiently homogeneous to omit additional B1 correction terms [33,34]. First, the flip angle distribution across the slice profile was calculated by simulating the Bloch equations for the radiofrequency excitation pulse [23,35]. This distribution was then used to generate a large dictionary of simulated signal recovery curves for a wide grid of known T1 (50-2000 ms) and T2 (10-500 ms) values. This dictionary serves as a lookup table that relates the apparent relaxation times derived from the fit to the true underlying relaxation times. The final, corrected T1 and T2 maps were then generated by applying this lookup table to the initially estimated maps using linear interpolation.

**2.6 Registration Performance Evaluation**

To quantitatively evaluate the performance of the retrospective motion correction framework, two independent metrics were assessed across the dynamic time series. First, to evaluate structural alignment in the presence of the severe, non-linear intensity transformations inherent to the inversion recovery curve, Normalized Mutual Information (NMI) [36,37] was calculated between each individual dynamic frame and the selected mid-phase reference target. Higher NMI values indicate a reduction in joint entropy, confirming closer statistical alignment of tissue boundaries across varying image contrasts.

Second, to translate registration performance into physical spatial dimensions, anatomical displacement was tracked using a Center of Mass (CoM) analysis. The superior-inferior CoM coordinates of the segmented lung parenchyma were calculated across the acquisition series. To eliminate variations caused by the changing signal intensity profiles of the inversion recovery curve, this tracking was restricted exclusively to the fully recovered steady-state frames at the end of each inversion block, where anatomical boundaries are sharpest and signal intensities are physically stable. Residual registration error was then quantified by first calculating the within-subject standard deviation of the CoM position across the dynamic blocks for each individual, and subsequently computing the mean of these standard deviations across the cohort, reflecting the physical sub-millimeter stability of the lung parenchyma during free-breathing.

### 2.7 Automated External Processing

Following data acquisition, images were automatically transferred to an external dedicated processing workstation for post-processing. The entire computational workflow was implemented in Python, leveraging CUDA for parallel processing of computationally intensive steps. Specifically, the lung and vessel segmentation, and the voxel-wise non-linear fitting processes were performed using CUDA acceleration on a dedicated GPU. Processing was executed on a high-performance workstation equipped with a dual AMD EPYC 7502 32-core processors (128 CPU threads), NVIDIA Quadro RTX 8000 (48 GB VRAM) GPU, and 256 GB of RAM. The total time required to process a comprehensive 10-slice dataset from raw image data to final quantitative maps is 50 seconds. Upon completion of the pipeline, the generated quantitative maps were automatically routed to the hospital's Picture Archiving and Communication System.

## 3. RESULTS

The quantitative accuracy of the proposed method was verified in a phantom (**Figure 2**). The generated maps exhibited high spatial homogeneity, and the measured relaxation times T1 = (742 ± 4) ms; T2 = (213 ± 14) ms demonstrated high concordance with the ground truth values of T1 = (736 ± 1) ms and T2 = (210 ± 10) ms as determined by a standard spin-echo protocol.

To experimentally evaluate the underlying thoracic B0 field distribution and validate the assumption of negligible off-resonance effects at 0.55T , in-vivo B0 field maps were assessed in a representative subset of the volunteer cohort. The resulting field distributions demonstrated excellent spatial homogeneity across the lung parenchyma, with mean off-resonance shifts of (16.3 ± 7.0) Hz (with an average voxel-wise standard deviation of 32.6 Hz) across the thoracic cavity (**Supplementary Material Figure 1**).

The proposed free-breathing relaxometry technique was applied in all subjects, producing high-quality quantitative T1 and T2 maps across the entire lung volume without noticeable motion artifacts. To assess the performance of the retrospective motion correction pipeline, a multi-slice spatial comparison between a conventional breath-hold IR-uf-bSSFP acquisition and the proposed free-breathing framework was performed across all acquired slices covering the entire lung volume for three healthy volunteers. While the quantitative evaluation utilized the complete multi-slice datasets, three representative slices are showcased in **Figure 3** to visually demonstrate registration performance across different anatomical levels. The resulting map quality and quantitative values demonstrated close parametric concordance between the two methods, yielding an overall mean absolute percentage difference across all evaluated subjects and slices of (3.8±0.9)% for T1 and (8.4±2.8)% for T2.

Quantitative registration analysis in the entire volunteer cohort demonstrated highly stable structural alignment across the dynamic inversion recovery time series. Following automated motion correction, the NMI between individual dynamic frames and the mid-phase reference target significantly increased from a baseline of 0.46±0.04 to 0.81±0.03 across the cohort, indicating robust motion elimination despite severe non-linear intensity variations. To translate this alignment into physical dimensions, tracking of the lung parenchyma's superior-inferior CoM was performed on high-contrast steady-state frames across the acquisition blocks. Prior to registration, respiratory motion introduced a mean bulk anatomical displacement of (12.4±2.8) mm across the cohort, which the automated pipeline successfully suppressed to a residual tracking error of (0.8±0.2) mm, confirming sub-millimeter structural stabilization during free-breathing.

Illustrative coronal T1 and T2 maps from four additional representative healthy volunteers, with and without pulmonary vessels excluded, are shown in **Figure 4** alongside voxel-wise goodness-of-fit ($R^2$) maps. The superimposed $R^2$ maps demonstrated spatial homogeneity and high mathematical model consistency throughout the segmented lung parenchyma (mean $R^2$ = 0.91±0.06), confirming that the automated motion-correction framework preserved the underlying transient bSSFP signal evolution curve without introducing regional fitting artifacts or model-related inaccuracies.

To illustrate the clinical applicability, **Figure 5** presents an exemplary slice from a 62-year-old male patient with confirmed Stage IIIA lung carcinoma and verified fibrosing hypersensitivity pneumonitis. CT imaging confirmed a progressive solid mass lesion in the apicoposterior segment of the left upper lobe. In the generated maps, this dense tumor tissue coincides spatially with distinctly elevated T1 relaxation times, distinguishing it from the surrounding parenchyma.

The reproducibility of the method was evaluated through ten measurements in one volunteer, where the subject was removed and repositioned between scans to mimic separate sessions and test scanner adjustment stability. These measurements showed high consistency in the derived relaxation times (**Figure 6**).

Quantitative results from the group of eight healthy volunteers and the patient are summarized in **Supplementary Material Table 1**. The final analysis was performed on the automatically segmented lung parenchyma, excluding pulmonary vessels. The average (± standard deviation) T1 and T2 values across all healthy subjects for the total lung parenchyma were (930±40) ms and (90±8) ms, respectively.

## 4. DISCUSSION

In this study, we introduced and validated a fully automated method for simultaneous T1 and T2 mapping of the lung at 0.55T. The primary components of this approach consist of a free-breathing acquisition strategy, eliminating the need for external triggering or gating hardware, and a corresponding automated post-processing pipeline. Together, these advancements demonstrate the feasibility of a free-breathing approach that could overcome primary barriers to the clinical adoption of quantitative lung MRI. We implemented the IR-uf-bSSFP [23] sequence at 0.55T to benefit from the overall reduced field inhomogeneity and thus possible off-resonance related quantification issues.

While 0.55T inherently presents lower SNR, we strategically employed a bSSFP sequence. Beyond providing superior signal-per-unit-time [21] which mitigates low SNR, the coherent nature of bSSFP is essential for the simultaneous T1 and T2 quantification. In contrast, incoherent methods (e.g., FLASH) which are typically limited to T1 mapping. Furthermore, the free-breathing acquisition with multi-repetition averaging allows for signal accumulation, effectively compensating for the lower intrinsic SNR at 0.55T [38]. Additionally, this approach provides a stable reference for motion correction, which aids in mitigating potential perfusion-related artifacts and cardiac pulsation effects without the strict requirement for ECG-gating.

We observed mean parenchymal T1 of (930±40) ms and T2 of (90±8) ms across eight healthy subjects. These results are consistent with the known field-strength dependence of relaxation times, where T1 shortens and T2 lengthens at lower fields [ref needed]. Additionally, the T2 of pulmonary vessels (160±10) ms was significantly higher than that of the parenchyma, reflecting the higher water content of blood (**Supplementary Material Table 1**).

Proof-of-concept application in a patient demonstrated that the derived quantitative maps could robustly differentiate pathological tissue (solid mass) from normal parenchyma. Specifically, the solid tumor mass coincided spatially with distinctly elevated T1 relaxation times, agreeing with the expected signal behavior of solid tissue [39,40]. The ability of this fully automated, free-breathing framework to capture such contrasts without user intervention highlights its clinical feasibility. This robustness could facilitate future studies investigating whether such mapping techniques might serve as complementary tools for characterizing pulmonary pathologies.

The elimination of breath-holding requirements broadens the applicability of the technique to patient populations unable to perform reliable breath-holds, including pediatric patients, elderly individuals, and those with severe respiratory compromise. Beyond improving accessibility, a free-breathing acquisition inherently simplifies the imaging workflow by removing the need for operator-guided breath-hold instructions and gating hardware, which may improve patient compliance and examination throughput in routine clinical settings.

Several limitations of this study should be acknowledged. A key technical constraint is that the current 2D multi-slice approach acquires each slice at a different point in time. Although motion is corrected effectively within each individual slice, slice-to-slice misalignments due to respiratory drift or slow body shifts over the full 3-minute acquisition period are still possible. Furthermore, while this study provides initial normative data, a larger study would be beneficial to establish a more definitive range of values. Finally, while we corrected for slice profile imperfections, other potential confounds, such as magnetization transfer effects or B1 field inhomogeneities, were not explicitly modeled in this work.

## 5. CONCLUSION

In conclusion, we have shown that simultaneous mapping of lung parenchymal T1 and T2 can be implemented at 0.55T using a free-breathing IR-uf-bSSFP acquisition. By eliminating the need for breath-holding and external gating hardware, this technique improves patient compliance and accessibility. Future work will focus on evaluating the diagnostic potential of this method across a broader range of pulmonary pathologies.

## 6. ACKNOWLEDGEMENTS

This work was supported by the Swiss National Science Foundation, Grant/Award Number: SNF 320030_219186.

# FIGURES

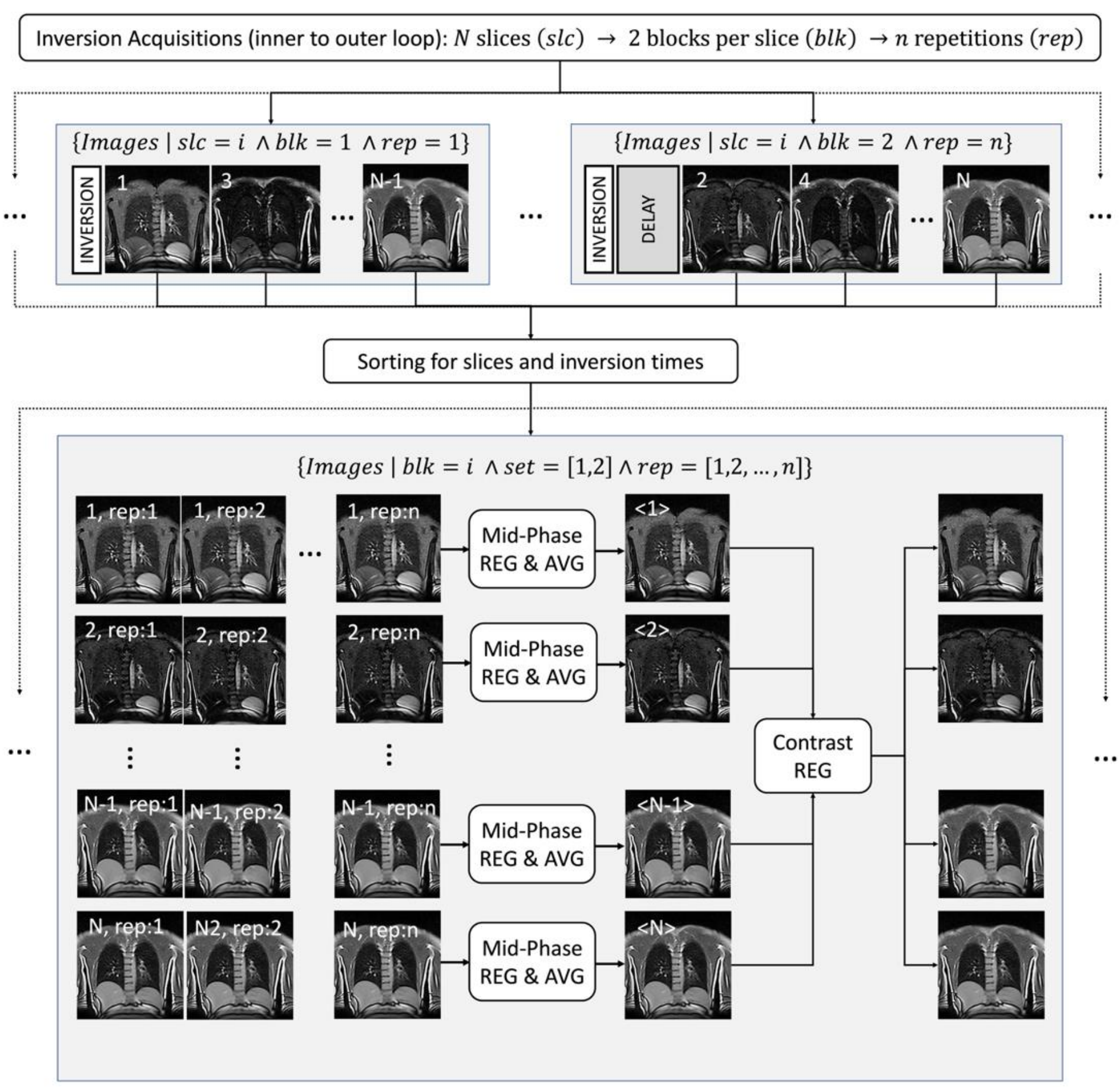

**Figure 1:** Schematic of the free-breathing data acquisition and automated image processing pipeline. The acquisition strategy utilizes two separate inversion blocks per slice to double the temporal sampling density of the signal recovery curve. A delay of half the acquisition time is applied in the second block (right), effectively interleaving the sampling time points (e.g., odd vs. even indices) relative to the inversion pulse. The two-step retrospective motion correction framework. First, repetitions acquired at the same inversion time point are co-registered to a mid-respiratory reference frame and averaged ("Mid-Phase Reg & Averaging") to minimize respiratory motion and improve SNR. Second, the resulting averaged images—which exhibit varying contrast across the recovery curve—are aligned using a symmetric diffeomorphic registration algorithm ("Contrast Registration") to ensure spatial consistency across the entire time series.

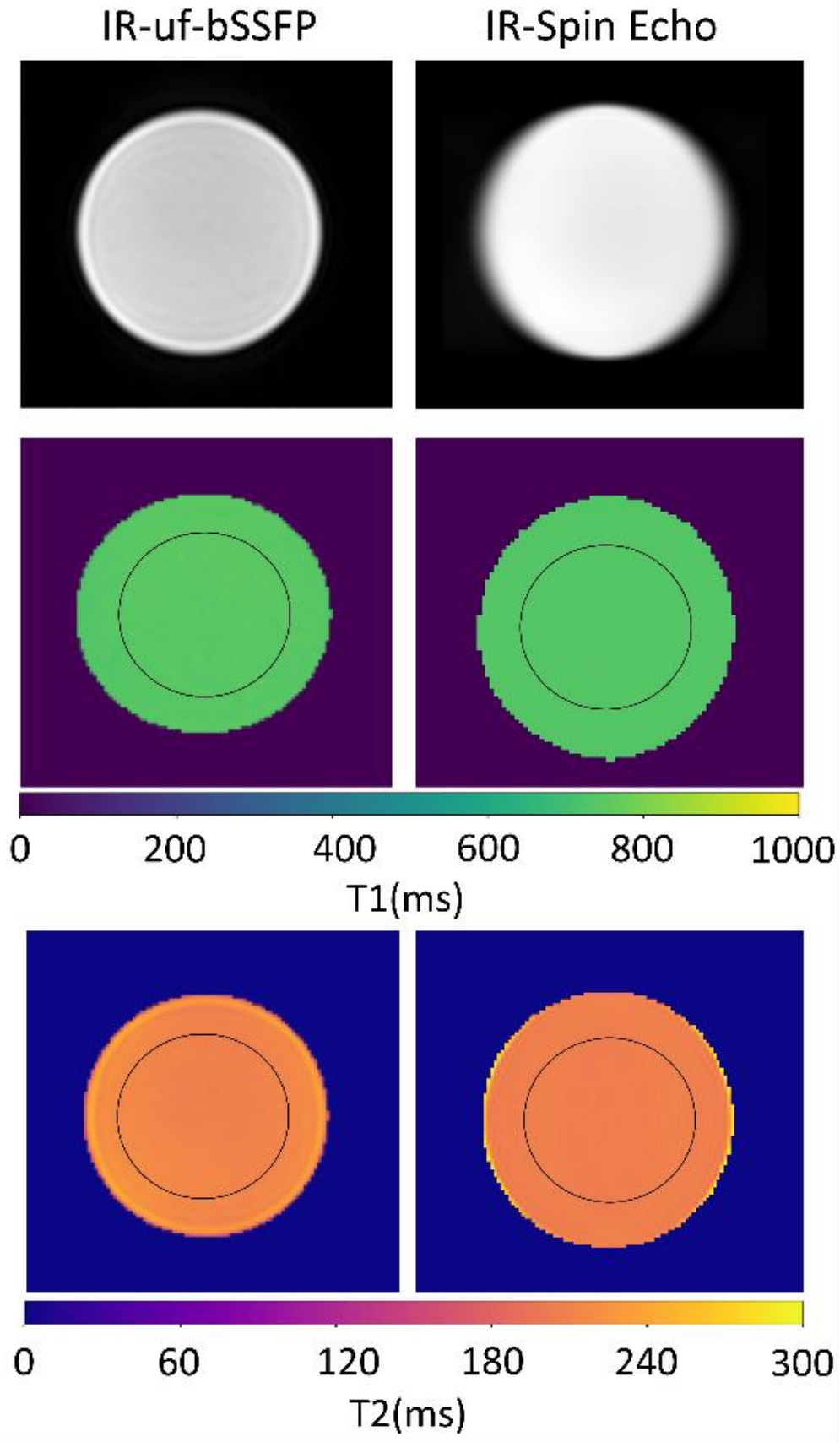

**Figure 2**: Validation of the quantitative accuracy of relaxometry in a structural phantom, where the left column displays parametric maps obtained using the proposed inversion-recovery ultra-fast balanced steady-state free precession (IR-uf-bSSFP) sequence and the right column shows the corresponding ground-truth reference maps derived from multi-point turbo spin-echo protocols. From top to bottom, the rows display representative magnitude images of the phantom structure, quantitative T1 maps, and quantitative T2 maps, respectively. Quantitative analysis within a central region of interest (ROI) demonstrated high parametric concordance between the proposed IR-uf-bSSFP sequence (T1 = 742 ± 4 ms; T2 = 213 ± 14 ms) and the reference standards (T1 = 736 ± 1 ms; T2 = 210 ± 10 ms). The generated parametric maps demonstrate excellent spatial homogeneity across all compartments, confirming the baseline accuracy and technical stability of the mapping framework prior to in-vivo translation.

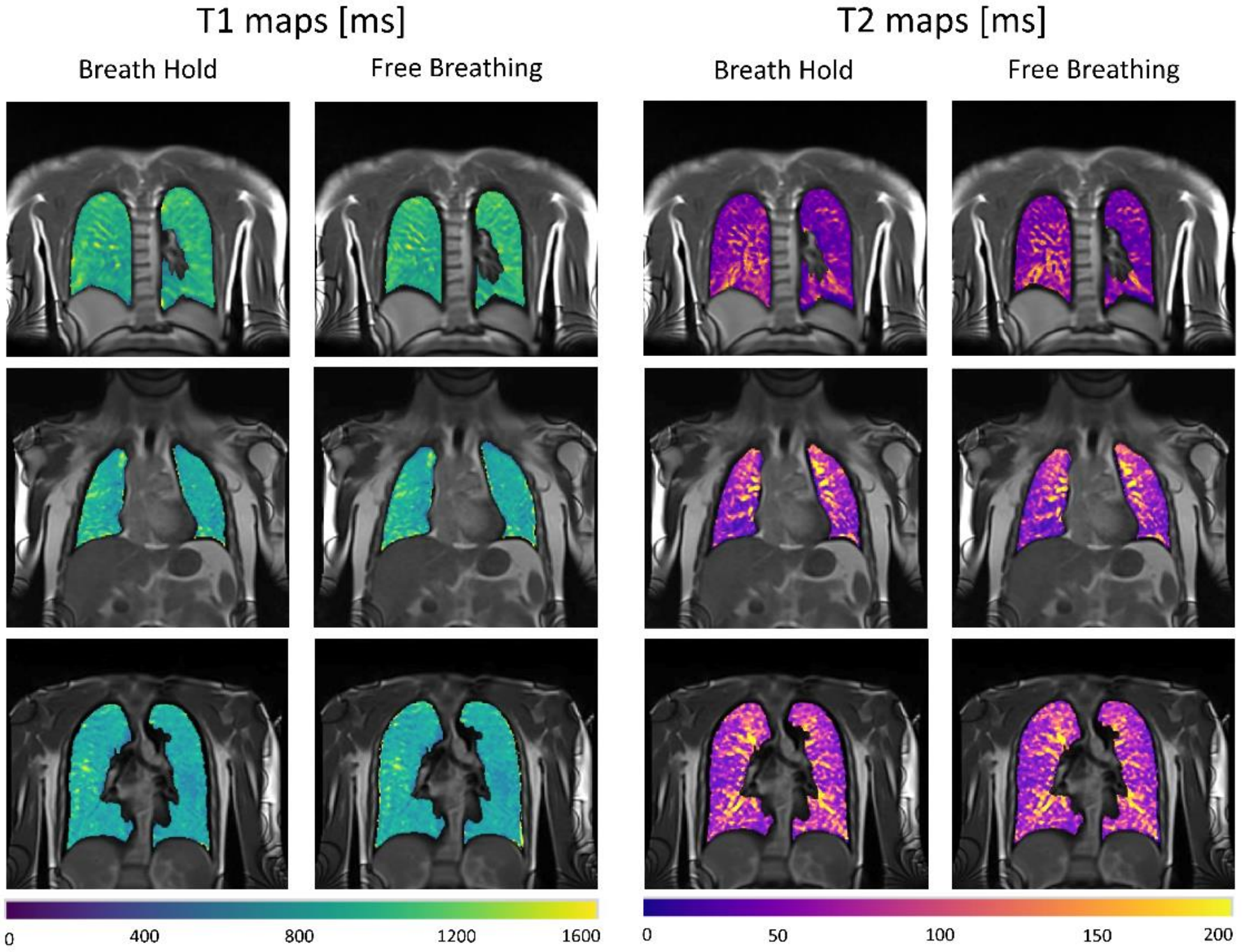

**Figure 3:** Comparison of quantitative maps between breath-hold (BH) and free-breathing (FB) acquisitions across three representative slices and volunteers. The panels display T1 maps (left columns) and T2 maps (right columns). For each slice level (rows), parametric maps acquired from healthy volunteers during a breath-hold at a mid-respiratory position are compared directly against corresponding maps acquired using the integrated free-breathing protocol with retrospective motion correction. To ensure an appropriate metric comparison and maximize structural alignment between the two frameworks, the static breath-hold images were utilized as the registration target for the free-breathing pipeline. The high spatial concordance and close visual agreement between the map pairs demonstrate that the automated motion correction pipeline yields quantitative relaxation times comparable to those obtained from a static breath-hold reference across all three anatomical layers.

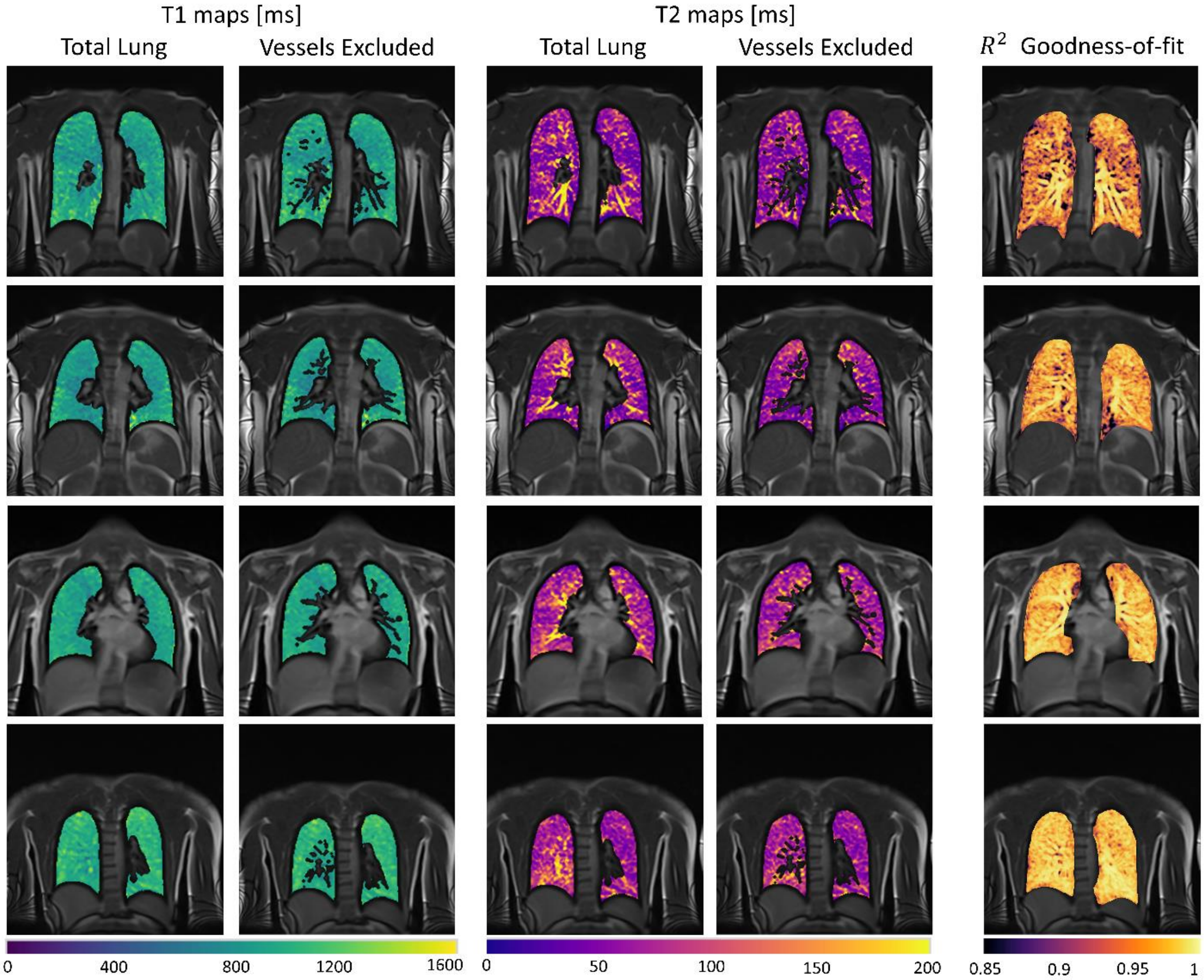

**Figure 4:** Representative quantitative maps obtained from four healthy volunteers using the proposed free-breathing framework. The panels display T1 maps, T2 maps, and voxel-wise coefficient of determination ($R^2$) maps. For each volunteer (rows), the T1 and T2 maps are presented both for the complete lung volume and for the lung parenchyma with major pulmonary vessels masked out. The cohort included two male subjects (top two rows; 24 and 30 years old) and two female subjects (bottom two rows; 25 and 26 years old). The consistent map quality and acceptable mathematical model fit observed across different subjects (mean parenchymal $R^2$ = 0.91 ± 0.06) suggests the feasibility of the integrated processing pipeline against anatomical and slice position variability.

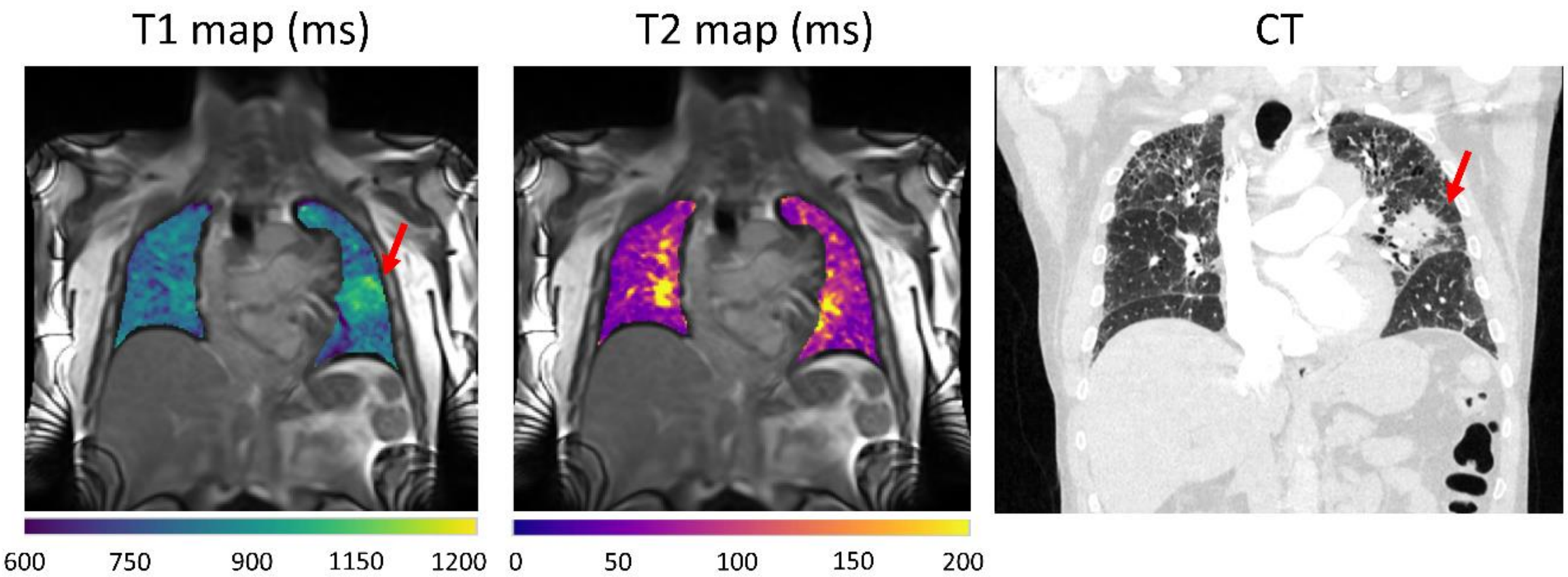

**Figure 5:** Clinical example of free-breathing quantitative mapping at 0.55T in a patient with confirmed pulmonary pathology. The subject is a 62-year-old male diagnosed with fibrosing interstitial pneumopathy (hypersensitivity pneumonitis) and suspected Stage IIIA lung carcinoma. (Left) The T1 relaxation map reveals distinctly elevated relaxation times corresponding to the dense tissue of the solid tumor (arrow, note that the color scale has been adjusted to enhance the visualization of the lesion contrast against the surrounding parenchyma). (Middle) The corresponding T2 map which does not clearly depict the tumor. (Right) Clinical reference CT confirms the presence of a solid mass lesion located in the apicoposterior segment of the left upper lobe (arrow), alongside background parenchymal changes consistent with fibrosis. The quantitative T1 signature in the left panel coincides spatially with the dense tumor tissue identified in the right panel (CT scan), distinguishing it from the surrounding aerated lung.

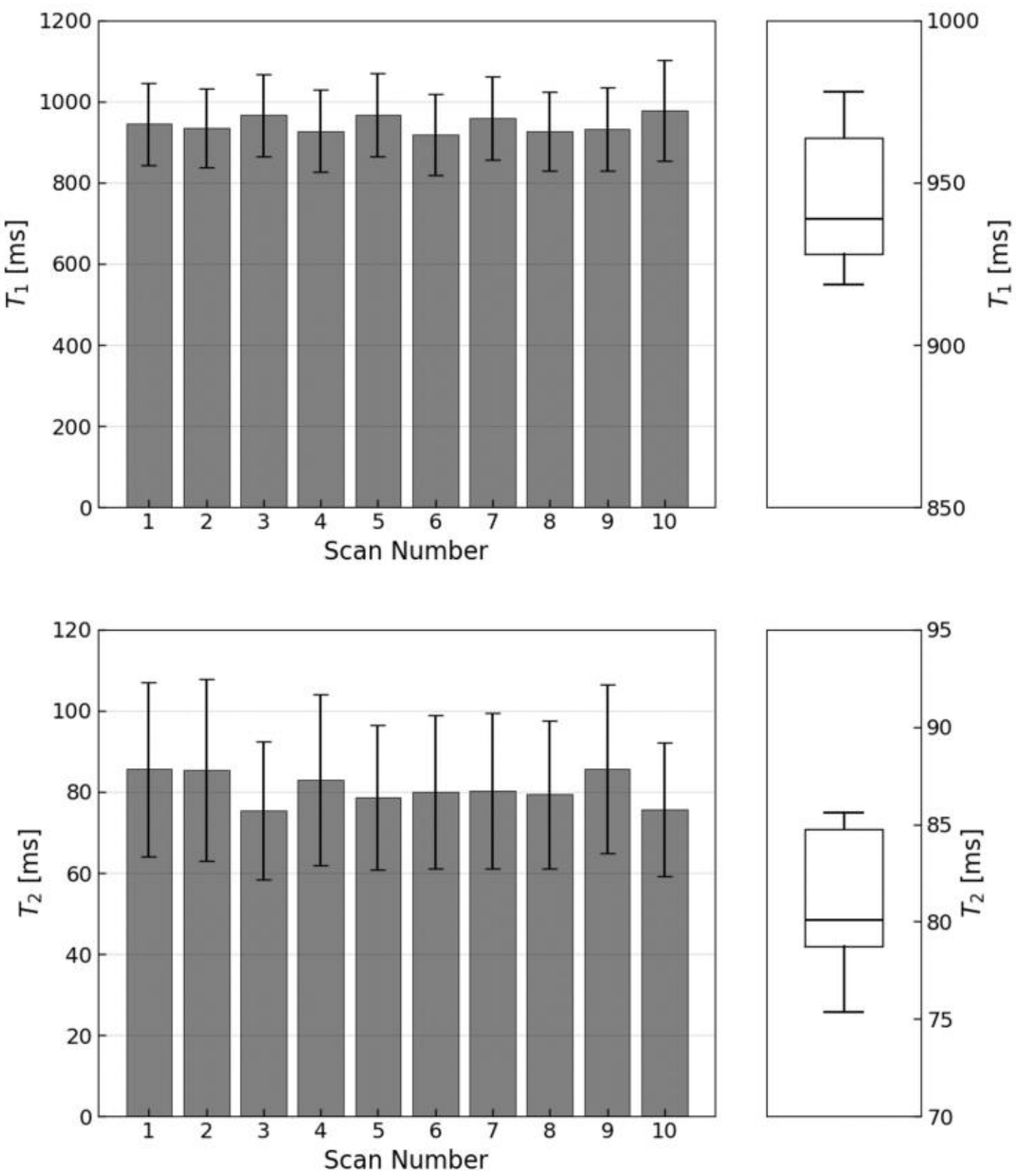

**Figure 6:** Reproducibility of Quantitative Lung T1 and T2 Measurements. The reproducibility of the automated mapping technique against positioning variability was assessed. The variability of (a) T1 and (b) T2 relaxation times is shown, measured in the segmented lung parenchyma of a single volunteer across 10 consecutive scanning sessions. For each session, the volunteer was removed from the scanner and then repositioned to test the inter-acquisition stability of the measurements. The bar plots show the mean and standard deviation for each of the 10 independent scans. The box plots on the right summarize the median (solid line), lower and upper quartile values (box ends), and the overall range (whiskers), visually confirming the high stability and clinical robustness of the derived quantitative parameters despite repositioning and anatomical variability.

**SUPPLEMENTARY MATERIAL FIGURES**

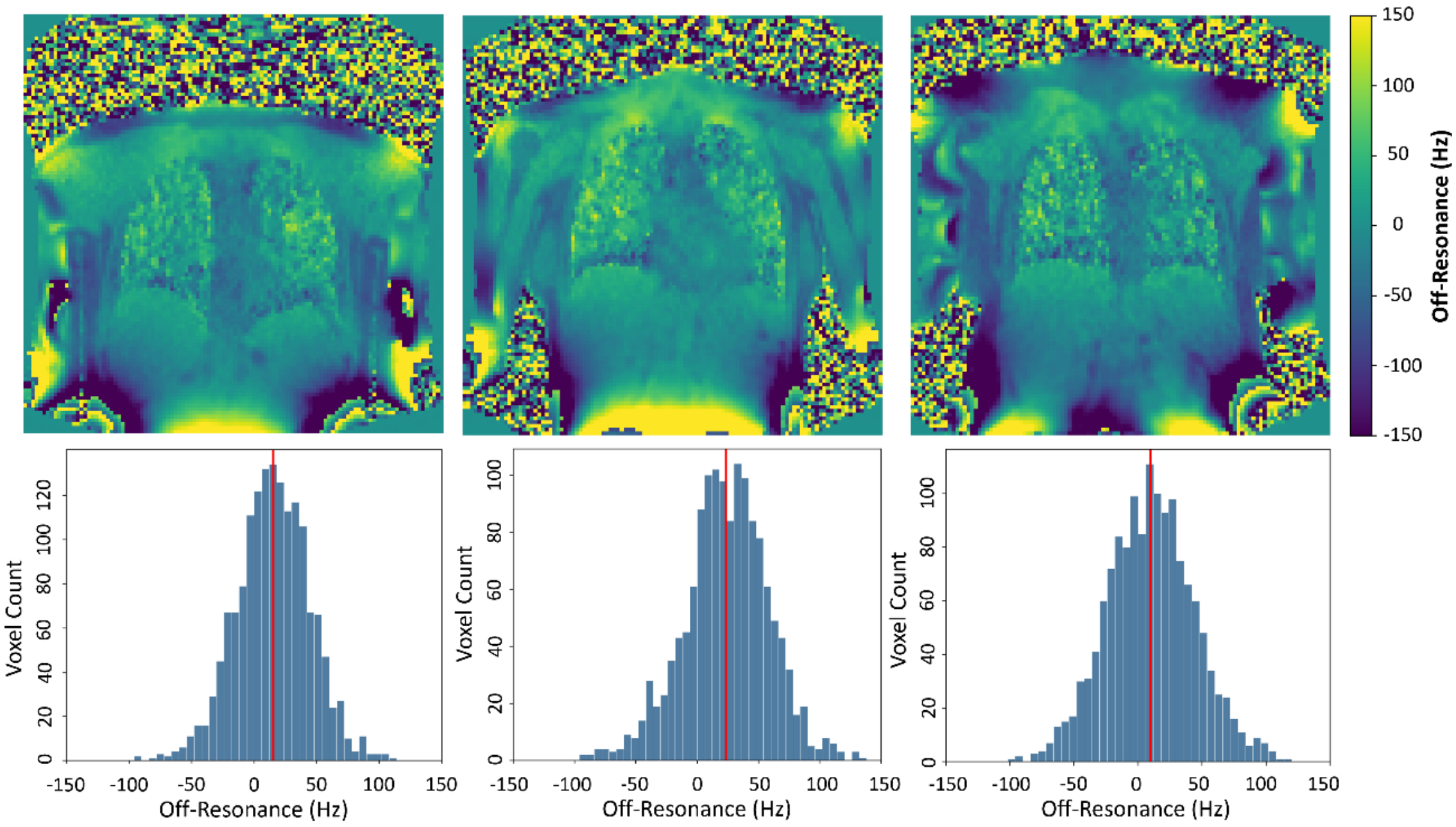

**Supplementary Material Figure 1**: Assessment of off-resonances in a representative subset of the healthy volunteer cohort (N = 3) at 0.55T. For the evaluated cohort, the observed parenchymal off-resonance distributions (visualized in the top row and quantified in the corresponding bottom-row histograms) were closely confined within ±100 Hz. This range falls completely within the flat passband profile of the transient bSSFP sequence—defined in literature as ±1/(3TR) [41] —which corresponds to ±111 Hz for the protocol's TR of 3.0 ms. Consequently, these minimal local field variations exert a negligible bias on the assumed on-resonance transient signal trajectory, validating the voxel-wise fitting model (see Equations 1–3).

| | Left Lung – Vessels Excluded (all slices) | | Right Lung – Vessels Excluded (all slices) | | Full Lung – Vessels Excluded (all slices) | | Lung Vessels (all slices) | |
|---|---|---|---|---|---|---|---|---|
| | T1 (ms) | T2 (ms) | T1 (ms) | T2 (ms) | T1 (ms) | T2 (ms) | T1 (ms) | T2 (ms) |
| Volunteer 1 (F) | 982 ± 92 | 89 ± 28 | 984 ± 93 | 83 ± 30 | 983 ± 93 | 86 ± 30 | 970 ± 70 | 145 ± 40 |
| Volunteer 2 (F) | 935 ± 78 | 91 ± 30 | 941 ± 72 | 83 ± 28 | 937 ± 79 | 88 ± 39 | 909 ± 68 | 153 ± 43 |
| Volunteer 3 (F) | 951 ± 114 | 89 ± 37 | 943 ± 105 | 81 ± 37 | 947 ± 110 | 85 ± 30 | 907 ± 75 | 164 ± 51 |
| Volunteer 4 (F) | 988 ± 115 | 87 ± 38 | 958 ± 104 | 85 ± 34 | 986 ± 109 | 87 ± 36 | 953 ± 88 | 160 ± 50 |
| Volunteer 5 (M) | 889 ± 124 | 89 ± 42 | 904 ± 112 | 89 ± 41 | 896 ± 120 | 89 ± 41 | 856 ± 80 | 170 ± 61 |
| Volunteer 6 (M) | 902 ± 88 | 105 ± 40 | 924 ± 86 | 101 ± 38 | 912 ± 93 | 103 ± 40 | 889 ± 55 | 173 ± 43 |
| Volunteer 7 (M) | 901 ± 120 | 78 ± 38 | 875 ± 103 | 82 ± 40 | 889 ± 114 | 80 ± 40 | 882 ± 79 | 161 ± 52 |
| Volunteer 8 (M) | 887 ± 106 | 91 ± 28 | 891 ± 96 | 89 ± 31 | 889 ± 101 | 87 ± 34 | 862 ± 81 | 182 ± 50 |
| Patient (M) | 845 ± 82 | 84 ± 34 | 815 ± 51 | 84 ± 27 | 828 ± 64 | 84 ± 27 | 870 ± 65 | 146 ± 46 |

**Supplementary Material Table 1**: Quantitative T1 and T2 relaxation times (mean ± standard deviation) for the eight healthy volunteers as well as the patient included in this study. Data are reported for the left and right lungs separately, the combined full lung volume, and the segmented pulmonary vessels. Note that for the lung parenchyma regions (Left, Right, and Full Lung), major pulmonary vessels were excluded to ensure the values reflect tissue relaxation properties rather than blood pool signal.